\documentclass{ringb99}
\def\h5{\h_{50}}
\def\beq{\begin{equation}}
\def\eeq{\end{equation}}
\def\eps{\varepsilon}
\def\invis#1{#1}
\begin{document}
\title{Radio Ghosts}
\author{T.A. En{\ss}lin\inst{1,2}}  
\institute{Department of Physics, University of Toronto, 60 St. George 
  Street, Toronto, ON, M5S1A7, Canada \and Max--Planck--Institut f\"ur
  Radioastronomie, Auf dem H\"ugel 69, 53121 Bonn, Germany} 
\maketitle

\begin{abstract}
  We investigate the possibility that patches of old radio plasma
  (`radio ghosts') of former radio galaxies form a second distinct
  phase of the inter-galactic medium (IGM), not mixed with the thermal
  gas.  The separation of this phase from the ambient gas and its
  resistance against eroding turbulent forces is given by magnetic
  fields, which are expected to be roughly in pressure equilibrium
  with the surrounding medium.  Since patches of this plasma are
  largely invisible in the radio we use the term `radio ghost' to
  characterize their nature.  Possibilities and difficulties of
  different detection strategies of ghosts are discussed. These
  involve radio emission, cosmic microwave background (CMB) and
  starlight Comptonization, and Faraday rotation. Re-activation of the
  electron population in shock waves of cosmological structure
  formation, which seems to lead to the cluster radio relic phenomena.
  We discuss the role radio ghosts can have: They are able to store
  relativistic particles for cosmological times, but are also able to
  release them under the influence of very strong turbulence.  This
  might happen during a major merger event of clusters of galaxies.
  The released relativistic proton population could produce the
  observed radio halos of some cluster of galaxies via hadronic
  reactions with the background gas leading to the production of
  secondary electrons and positrons.

  Destroyed ghosts, mixed with the IGM can help to magnetize it. Finally,
  the strong field strength within ghosts should have a significant
  impact on the propagation of extragalactic high energy cosmic rays.
\end{abstract}

\section{Introduction}

Active radio galaxies fill large volumes in the IGM with radio plasma.
Over cosmological time-scales this plasma becomes rapidly invisible to
radio telescopes due to inverse Compton (IC) and synchrotron energy
losses of the relativistic electrons. Afterwards it might form an
invisible, but possibly important phase of the IGM. The amount of
energy stored in this form in the IGM is difficult to estimate.
Various approaches have been made, depending on different assumptions,
but giving comparable results (En{\ss}lin et al.  1997, 1998a). The
simplest is to assume that the power of active galactic nuclei
deposited into radio plasma and into X-ray light output are
comparable.  The integrated energy density (per co-moving volume) in
quasar light emitted during (and estimated for) the epoch of quasar
activity is $(0.9 - 1.3)\cdot 10^{-15}\, {\rm erg\, cm^{-3}}$ (Chokshi
\& Turner 1992). The energy density of the radio plasma should be the
same, if our assumption holds. For comparison: if this energy were to
be thermalized and transferred to the gas in the Universe, it would
give a heat input of $kT = (1.3 - 1.9)\; {\rm keV}$ $(\Omega_{\rm b}
h_{50}^{2} /0.05\,)^{-1}$ at $z \approx 2$ (assuming here and in the
following $H_{\rm o} = 50\, h_{\rm 50}\,\,{\rm km\,s^{-1}\,Mpc^{-1}}$)
.

But we argue that this does not necessarily happen. The radio plasma
and later radio ghosts will expand or contract until they reach
pressure equilibrium with the surrounding medium. The pressure of the
ghost is given by that of the confined relativistic particles and the
magnetic fields, assumed to be in rough energy equipartition.
Therefore magnetic fields should be typically of the strength of the
thermal energy density of the environment.  Subsonic turbulence in
this environment, which has an energy density below the thermal energy
density, is therefore not strong enough to overcome the magnetic
elastic forces of the radio ghost. Sonic or super-sonic turbulence,
which is e.g. expected in giant merger events of cluster of galaxies,
can `shred' the ghost into smaller pieces. The size of such pieces
will be comparable to the eddy size of the turbulence. This means,
since a typical turbulent spectrum has less energy density on smaller
scales, that there is a length-scale below which the turbulence is not
able to overcome magnetic forces.  Turbulent erosion of radio ghosts
should stop at this length-scale, leaving small-scale patches of still
unmixed old radio plasma.

\section{Detection of Radio Ghosts}
The number density, sizes, and filling-factor of our hypothesized
radio ghosts is poorly constrained due to the uncertainties in the
above considerations. But this knowledge would be required in order to
estimate their influence on the properties of the IGM.  Observational
evidence of their existence and measurements of their properties would
therefore be of great value. But as we show in the following, this is
hard to achieve, justifying the name `radio ghosts'.
\subsection{Synchrotron Emission}
The old population of relativistic electrons within the ghost is
emitting low frequency radio emission. At a given frequency $\nu$
synchrotron emission reveals mostly electrons with a Lorentz-factor of 
\begin{equation}
  \gamma(\nu,B,z) = \sqrt{ 2 \pi\; m_{\rm e}\; c\; \nu\; (1+z)\; /(3\,
    e\, B)}\;,
\end{equation}
where $B$ is the magnetic field strength and z the redshift of the
emission region.  Electrons will cool due to synchrotron and IC
losses with a cooling time of 
\begin{equation} 
  t_{\rm cool} (\gamma, B, z) = \left(\frac{4}{3}
    \frac{\sigma_T}{m_{\rm e} \,c}\, (\eps_B + \eps_{\rm CMB})\,
    \gamma \right)^{-1}\;,
\end{equation}
where $\eps_B = B^2/(8 \pi)$ and $\eps_{\rm CMB} = \eps_{\rm CMB,o}
(1+z)^4$ are the magnetic field and CMB energy density, the latter
expressed by its present day value $\eps_{\rm CMB,o}$.  The cooling
time of electrons visible at a frequency $\nu$ is $t_{\rm
  cool}(\gamma(\nu,B,z), B, z)$. This function has a maximum at
$\eps_B = \eps_{\rm CMB}/3$ for fixed $\nu$ and $z$. Therefore any
electron visible at a given frequency $\nu$ had to be accelerated no
later than
\begin{equation}\label{eq:tcoolmax}
  t_{\rm cool,max}(\nu ,z) = 0.7\, {\rm Gyr}\; (\nu/{100\, \rm
    MHz})^{-1/2} \, (1+z)^{-7/2}
\end{equation}
ago, otherwise it could not have retained its energy. This is an upper
limit, and in reality the lifetime of radio visible electrons is
always shorter: if the magnetic fields are strong ($\eps_B >
\eps_{CMB}/3$) synchrotron cooling is stronger, if the magnetic fields
are weak ($\eps_B < \eps_{CMB}/3$) high energy electrons are visible
at the observing frequency $\nu$ which also have shorter IC-cooling
times than given by Eq.  \ref{eq:tcoolmax}.  Thus, low frequency radio
observation cannot reveal ghosts which are older than about a Gyr,
unless the electron population has been recently re-accelerated.

\subsection{CMB Comptonization}

The relativistic electrons, if still present within ghosts, will scatter
the CMB photons to higher energies. Since the infrared and the optical
bands are overwhelmed by other sources, the chance of detection only
exists above the UV range. In order to have a typical energy of 10 eV
for a scattered CMB photon $\gamma \approx 100$ is required, since
$<\eps_{\rm IC}> = \frac{4}{3}\, \gamma^2\, 2.7 \, kT_{\rm CMB}$
(Blumenthal \& Gould 1970). Thus, it has to be examined if energetic
electrons can maintain an energy of 50 MeV for cosmological times.

Again we want to examine the most optimistic case in order to
demonstrate the difficulties of the detection of ghosts. The minimal
cooling is given by the effect of CMB-IC cooling alone. Pure CMB-IC-cooling 
of electrons is governed by
\begin{equation} 
  \frac{d\gamma}{dt} = - \frac{\gamma}{t_{\rm cool}(\gamma, 0, z)} = -
  a_{\rm IC,o} \, \gamma^2\, (1+z)^4\;,
\end{equation}
with $a_{\rm IC,o} = \frac{4}{3}\, \sigma_T\, \eps_{\rm CMB,o}
/(m_{\rm e}\,c)$. Assuming for simplicity an Einstein-de-Sitter
Universe\footnote{The cosmic time is given by $t(z) = 2/(3\,H_{\rm o})
  (1 - (1+z)^{-3/2})$.} one finds that for an arbitrary high injection
energy at $z_{\rm inj}$ every electron has cooled below
\begin{equation}
  \gamma_{\rm max} = \frac{5 \, H_{\rm o}}{2\, a_{\rm IC,o}}
  \frac{1}{(1+z_{\rm inj})^{5/2} -1} \approx \frac{48\, h_{50}}{
    (1+z_{\rm inj})^{5/2} -1}\;.
\end{equation}
In order to get IC scattered CMB photons above 10 eV the last
acceleration of the electron population must have been no later than
$z_{\rm inj} = 0.16$, corresponding to an age of 4 Gyr. It is
therefore impossible to detect old radio ghosts from the cosmic epoch
of violent quasar activity around $z\approx 1...3$ by their IC-flux.
Synchrotron and other losses, which we neglected here, make the
detection only more difficult.


Another strategy might be to look for characteristic CMB decrements.
The relative number of up-scattered CMB photons is given by the
Thomson optical depth 
\begin{eqnarray}
  \nonumber
  \frac{\Delta I_{\rm CMB,ghost}}{I_{\rm CMB}} &=& \tau_{\rm ghost} =
  \sigma_T\, n_{\rm e, ghost} \, l_{\rm ghost} \\ &=& 2\cdot 10^{-6}
  \,\left( \frac{n_{\rm e,ghost}}{10^{-5}\,{\rm cm^{-3}}} \right)
  \,\left( \frac{l_{\rm ghost}}{100\,{\rm kpc}}\right)
\end{eqnarray}
A density of the relativistic electrons of $n_{\rm e, ghost} =
10^{-5}\,{\rm cm^{-3}}$ is optimistic but might apply to ghosts
confined in the dense ICM. Ghosts outside clusters of galaxies will
have relativistic electron densities which are lower by orders of
magnitude. Therefore the only chance to detect ghosts is inside
clusters of galaxies. And there, due to their low optical depth, only
a cluster averaged effect can produce a marginal detection: we assume
that a small fraction $\Phi_{\rm Vol}$ of the cluster volume is filled
with this remnant radio plasma. Further we assume pressure equilibrium
between the ghosts and the ICM, equipartition between magnetic fields
and relativistic electrons inside the ghosts\footnote{This is again
  optimistic, since a substantial fraction of the relativistic
  particle pressure might be given by relativistic protons.}, and a
typical relativistic electron energy of 1 MeV. For a 10 keV cluster
this gives a ratio of ghost- to ICM-optical depth of
\begin{equation}
  \frac{\tau_{\rm ghost}}{\tau_{\rm ICM}} \approx 0.03 \,\Phi_{\rm Vol}\;,
\end{equation}
determined by the ratio of relativistic to thermal electron number.
Photons scattered by relativistic electrons are practically removed
from the blackbody spectrum, whereas thermal electrons are only
shifted, leading to a decrement\footnote{This estimate is based on the
  maximal relative Sunyaev-Zeldovich decrement in flux at a frequency
  of $\nu \approx 1.8 \,kT_{\rm CMB}/h$ and a cluster temperature of
  10 keV.  We used the non-relativistic Kompaneets approximation,
  which should be sufficiently accurate for an order of magnitude
  estimate.} of
\begin{equation}
\frac{\Delta I_{\rm CMB,ICM}}{I_{\rm CMB}} \approx 0.06\, \tau_{ICM}\;.
\end{equation}
This implies that the relative strength of the ghost- and ICM-induced
CMB decrements is
\begin{equation}
  \label{eq:ic.dec.rel}
  \frac{\Delta I_{\rm CMB,ghost}}{\Delta I_{\rm CMB,ICM}} \approx 0.5\,
  \Phi_{\rm Vol} \ll 1\;.
\end{equation}
Ghosts are therefore hard to detect as a CMB decrement, but it can not
be excluded that future instruments will see them, if our optimistic
assumptions hold.
\subsection{Starlight Comptonization}
A low energy population of relativistic electrons, which may be
present in radio ghosts, is able to scatter the starlight to higher
energies. This was used as one possible scenario to explain the
extreme ultraviolet (EUV) excess of the Coma cluster of galaxies
(En{\ss}lin et al. 1999).  In this scenario the relativistic electrons
must be located within magnetic fields which separate them from the
dense ICM, otherwise high Coulomb losses would rapidly cool the
electrons. This would be the case for an old electron population
within radio ghosts.  A difficulty of this scenario is the relatively
high energy density required in the electron population to produce the
detected EUV radiation.  But if future observations with higher
angular resolution show that the EUV excess is clumped around the
cluster galaxies where the optical photon density is highest, this
would be an indicator for IC scattering by a strong low energy
electron population, which needs an environment as given within radio
ghost.
\subsection{Faraday Rotation}
Magnetic fields in radio ghosts are expected to be in rough
equipartition with the environment. Therefore one expects fields
strengths of $0.1....1\, \mu $G for ghosts outside, and up to a few
$10\,\mu $G for ghosts inside clusters of galaxies.  Fields of such
strength seem to be easily detectable at a first glance, since
magnetic fields of $\mu $G strength were seen in clusters of galaxies
by Faraday rotation measurements of linearly polarized background
radio sources (for a review: Kronberg 1994).  But Faraday rotation by
radio ghosts should be very difficult to detect for two reasons:
First, the electron number density is expected to be very low and
Faraday rotation measures the product of magnetic fields strength
times electron density.  Second, the electrons are expected to be
relativistic, and therefore Faraday rotation is suppressed (e.g.
Jones \& Hardee 1979). Ghosts can only become detectable by Faraday
rotation measurements, if the surrounding thermal gas is able to break
into the ghost.
\subsection{Cluster Radio Relics}
Radio ghosts are practically invisible as long as their electron
population remains at low energies. But if the population is
re-accelerated the ghost becomes radio luminous again. This can happen
when the ghost is dragged into a large-scale shock wave, e.g. in a
merger event of clusters of galaxies or at the accretion shock where the
matter is falling onto a cluster. The emission region is
expected to be irregularly shaped, and should exhibit linear
polarization due to the compression of the magnetic fields in the
shock. Such regions are indeed observed at peripheral locations of a
few clusters of galaxies and are called `cluster radio relics'. Their
properties, such as as degree and direction of polarization, surface
luminosity, peripheral position etc., can be understood within such a
scenario (En{\ss}lin et al.  1998b).  The observed rarity of the
cluster radio relic phenomena can be understood in this context: the
presence of a shock wave, which should be quite frequently found in
and at clusters of galaxies, is not sufficient to produce a cluster
radio relic, a radio ghost has to be present at the same location,
too.

Therefore cluster radio relics are up to now the best signature of the
existence of radio ghosts. On the other hand,  radio ghosts seem to be
sensitive tracers of large-scale shock waves and allow the study
of cosmological structure formation in action.
\section{The Possible Roles of Radio Ghosts}
\subsection{Storage Sites of Relativistics Protons}
Radio ghosts consist of magnetic fields and low energy relativistic
electrons.  A long outstanding question is the extent to which radio
plasma clouds contain relativistic protons. The detection of TeV
$\gamma$-rays from blazars support the presence of a significant
proton component, since this emission is difficult to understand
within pure leptonic jet models (Mannheim 1998)\nocite{mannheim98}. 

A relativistic proton population can therefore be expected within
ghosts, since the escape of protons is suppressed due to the low cross
field diffusion coefficient.  In order to get an idea of the relevant
diffusion time-scales we adopt a simplified version of the compound
diffusion described in Duffy et al. (1995). We estimate the escape
time of a 10 GeV proton in a 10 $\mu$G field, which could be typical
for a ghost in a cluster environment. The parallel and cross-field
diffusion coefficients
\begin{eqnarray}
  \label{eq:kappaperp}
  \kappa_\| &\approx& \frac{1}{3} \,c\,r_{\rm g}\,/\delta_{B}(r_{\rm g})\\
  \kappa_\perp &\approx& \frac{1}{3} \,c\,r_{\rm g}\,\delta_{B}(r_{\rm g})
\end{eqnarray}
depend on the (relative) energy density in magnetic fluctuation
$\delta_{B}(r_{\rm g}) = \delta B^2 (r_{\rm g})/B^2$ on the scale of
the gyro-radius $r_{\rm g} = 10^{-6}$ pc of the diffusing particle.
The magnetic fields of the ghost are in rough equipartition with the
surrounding thermal medium $\eps_B \approx \eps_{\rm th}$.  The level
of turbulence on the turbulence injection scale $l_{\rm inj} = 10$ kpc
is assumed to be a small fraction of the thermal energy density. We
adopt a value for the turbulence energy density integrated up to this
scale: $\eps_{\rm turb}(l_{\rm inj}) = 0.01 \,\eps_{\rm th}$.
Assuming a Kolmogoroff turbulence spectrum, one gets a turbulent
energy density integrated up to the scale $r_{\rm g}$, of
\begin{equation}
  \label{eq:smallturb}
  \eps_{\rm turb}(r_{\rm g}) = \eps_{\rm turb}(l_{\rm inj})\,
  (l_{\rm inj}/r_{\rm g})^{-2/3}\, \approx 10^{-8.7}\, \eps_{\rm
    th}\,.
\end{equation}
The turbulence-induced small-scale magnetic irregularities are
therefore $\delta_{B} \approx 10^{-8.7}$, leading to a cross field
diffusion coefficient of $\kappa_\perp \approx 10^{13.8}\,{\rm
  cm^2\,s^{-1}}$. This is far too small to allow any macroscopic
diffusion.

But the high parallel diffusion coefficient of $\kappa_\| \approx
10^{31.2}\, {\rm cm^2\, s^{-1}}$ allows the particle to travel rapidly
along the magnetic field lines. Now, if we allow neighboring field
lines to diverge exponentially, a small diffusive step of the particle
perpendicular to the field can be strongly amplified by the rapid
movement along the field. This leads to the compound cross-field
diffusion coefficient 
\begin{equation}
  \label{eqcompound}
  \kappa_{\rm comp} \approx \kappa_\perp (1+ \Lambda^2/\ln
  \Lambda)\approx 10^{25.6}\, {\rm cm^2\,s^{-1}}
\end{equation}
(Duffy et al. 1995), where the quantity
\begin{equation}
  \label{eq:Lambda}
  \Lambda = \frac{\delta_{B}(l_{\rm B})}{\sqrt{2}\,\delta_{B}(r_{\rm
      g})} \approx 10^{6.5}
\end{equation}
is simplified here by assuming a single typical field
correlation-length parallel and perpendicular to the field direction
of $l_{\rm B} \approx 10$ kpc. This is still too small to allow an
efficient escape of protons from the ghost, since a  typical diffusion time-scale is
\begin{equation}
  \label{esc1}
  \tau_{\rm diff} \approx (10 {\rm \, kpc})^2/(2 \,\kappa_{\rm comp})
  \approx 3\cdot 10^{11} \,{\rm yr}\,.
\end{equation}
On this argument, it seems reasonable that relativistic particles are
confined for cosmological times within ghosts.

\subsection{Escape of Relativistics Protons
\label{sec:release}}
Particles could escape if the diffusion coefficient is much higher.
This could be the case for a higher turbulence level than assumed
above.  During a major cluster merger the turbulence can get into
rough equipartition with the thermal energy density, thus being higher
by two orders of magnitude than assumed above. If the turbulence is
fully developed, the cross field diffusion coefficient $\kappa_\perp$
goes up by two orders of magnitude. $\Lambda$ should remain constant,
since it is mainly the ratio of large-scale to small-scale turbulent
energy, which does not depend on the normalization of the turbulence
energy spectrum. Therefore $\kappa_{\rm comp}$ increases also by a
factor of 100, giving a diffusion time of 3 Gyr, which is still large.

But turbulence needs some time to develop fully from large injection
scales to small dissipation scales. In a sudden merger the level of
large-scale turbulence will increase for some time, whereas the
small-scale turbulence is still low. The time-scale for a decay of a
turbulent eddy of scale $l$ with velocity $v_{\rm turb}$ is
approximately the eddy-turnover time $ l/v_{\rm turb}$. A Kolmogoroff
cascade needs therefore $\tau_{\rm turb} \approx \frac{3}{2} \,l_{\rm
  inj}/v_{\rm turb}(l_{\rm inj})$ to fully develop. Inserting $l_{\rm
  inj} = 10$ kpc, and $v_{\rm turb}(l_{\rm inj}) = 1000$ km/s for
strong turbulence in a galaxy cluster, one gets $\tau_{\rm turb}
\approx 15$ Myr.

During this period $\Lambda$ can be increased by a factor of 100. This
keeps $\kappa_\perp$ constant, but raises the compound diffusion
coefficient by 4 orders of magnitude. The resulting escape time of
$\tau_{\rm diff} \approx 30$ Myr is comparable to $\tau_{\rm turb}$
and allows therefore a large fraction of the confined protons to
escape\footnote{Both time-scales depend in a similar way on the chosen
  turbulence injection scale: $\tau_{\rm diff} \sim l_{\rm inj}^{2/3}$
  and $\tau_{\rm turb} \sim l_{\rm inj}$. Therefore a probably larger
  $l_{\rm inj}$ increases the ratio of $\tau_{\rm turb}$ and
  $\tau_{\rm diff}$, thus allowing more particles to escape within the
  period of strongly increased diffusion coefficient.}.  In this way a
strong cluster merger event can lead to a sudden injection of
relativistic protons into the dense ICM gas.  The protons may stay for
a cosmological time within the ICM, as simple diffusion-time estimates
suggest (En{\ss}lin et al. 1997; Berezinsky et al. 1997).

\subsection{Radio Halos}
Cluster radio halos are rare occurrences, as are cluster radio relics.
Their presence seem to be connected with a recent or on-going merger
event of the cluster (De Young 1992), although there are many clusters
showing evidence for strong merging there are very few radio halos
observed.

The models for cluster radio halo formation can divided into two
groups: primary and secondary electron models, which also could be
named leptonic and hadronic models. In leptonic models the radio
emitting electrons are accelerated or injected into the ICM directly
(Jaffe 1977; Rephaeli 1977, 1979; Roland 1981; Schlickeiser et al.
1987).  In hadronic models (Dennison 1980) protons are accelerated or
injected, and they interact hadronically with the ICM gas producing
charged pions which decay into electrons and positrons mainly by the
following reactions:
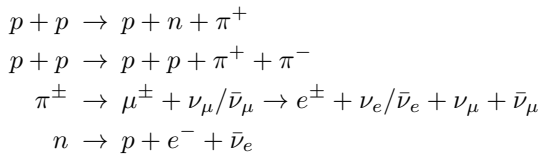
\begin{eqnarray}
  \label{eq:pp}\nonumber
  p + p &\rightarrow& p + n + \pi^+ \\
  p + p &\rightarrow& p + p + \pi^+ + \pi^- \nonumber\\
  \pi^\pm &\rightarrow& \mu^\pm + \nu_{\mu}/\bar{\nu}_{\mu} \rightarrow
  e^\pm + \nu_{e}/\bar{\nu}_{e} + \nu_{\mu} + \bar{\nu}_{\mu}\nonumber\\
  n &\rightarrow& p + e^- + \bar{\nu}_e\nonumber
\end{eqnarray}
The same number of electrons and pions are produced, but since more
$\pi^+$ than $\pi^-$ are generated, and the decay of a $\pi^+$ gives
more kinetic energy to the positron than the $\beta$-decay of a
neutron to the electron, the secondary positron spectrum dominates
over the secondary electrons at energies higher than the pion rest
energy. If radio halos have a hadronic origin, they are mainly
synchrotron emission from positrons.

The advantage of the hadronic model of halo formation is that the
lifetime of energetic protons in the ICM is of the order of a Hubble
time (En{\ss}lin et al. 1997), whereas radio emitting (at $\nu \approx
$GHz) relativistic electrons cool in less than 0.2 Gyr (see Eq.
\ref{eq:tcoolmax}). In the hadronic case this allows the protons to
diffuse over the large observed radio emitting volume, even if they
are released by point sources. In the leptonic models in-situ
acceleration has to take place since the cooling time of the
relativistic electrons is not sufficient to allow a travel distance to
be comparable to the radio halo size. This makes acceleration in a
cluster merger shock unlikely to be the explanation in the leptonic
framework, since the shock crossing time is of the order of a Gyr,
leading to a less symmetric emission profile than observed in
clusters.  On the other hand, in a hadronic model, where the protons
are injected during the shock passage, the low interaction rate of the
protons tend to smear out the effect of the time dependent shock
passage and permit the cluster to remain radio luminous after the
shocks disappeared.  If radio halos are really connected to merger
events, hadronic models are favored to explain their smooth observed
profiles.

We thus propose that proton release from ghosts during a cluster
merger event (Sec. \ref{sec:release}) is an attractive injection
mechanism for the hadronic halo formation scenario for several
reasons. First it would explain the connection between merger events
and radio halo formation. The protons are released in strong mergers
only, because the diffusion coefficient depends strongly on the level of
large-scale turbulence. Fermi acceleration in the turbulence or in
cluster shocks may further energize this relativistic population.
Note, that particle injection from a thermal distribution into a Fermi
acceleration mechanism is ineffective, but proton release from ghosts
would avoid this difficulty.

Simple shock acceleration models (without injection from ghosts) also
predict a higher proton acceleration rate in strong mergers. But the
rarity of cluster radio halos is difficult to understand in pure shock
acceleration scenarios, since we know of several strong merging or
post-merging clusters without a radio halo.  A good example is
Abell 3667, which shows evidence for a merger in X-rays (Markevitch et
al. 1998), in the galaxy distribution, and by the existence of two
radio relics (R\"ottgering et al.  1997). The morphology and location
of these relics was recently modeled in a hydrodynamic simulation with
simultaneous particle acceleration by Roettiger (1999). This cluster
is apparently in a post-shock situation, comparable to Coma (Burns et
al. 1995), but does not show any radio halo as Coma does.

The proposed ghost proton release scenario can give an explanation for
this, since nearly all the relativistic protons should escape from the
ghosts during one strong merger event. A second strong merger would
not have the same effect, since there are no further protons to
`release' after the first strong merger event. If this is the
explanation of the rarity of radio halos, it implies that many cluster
might have had their halos in the past, and therefore many more radio
halos should exist at higher redshifts than today.
\subsection{Magnetization of the IGM}

Magnetic tension should prevent the ghosts from being mixed with the
IGM. But it is difficult to state to what level ghosts really can
resist the turbulence, especially strong turbulence during major
cluster merger event. If ghosts become mixed with the IGM, they
release their relativistic particles\footnote{If thermal gas is able
  to get into the strong magnetic fields of a ghost, relativistic
  particles can easily escape.}, and magnetize the thermal IGM gas.
This can be a substantial source for the magnetic fields observed in
clusters of galaxies. IGM magnetization by radio galaxies, which is a
closely related idea, was proposed by Daly \& Loeb (1990).

\subsection{Propagation of Ultra-High Energy Cosmic Rays}
It is widely believed that the origin of ultra-high energy cosmic rays
(UHECRs) is extragalactic, since the galactic magnetic fields are not
strong enough to confine them, so that a extremely strong and
isotropically distributed population of galactic UHECR sources would
be required to fit the flux and distribution of arrival directions.

The travel length $L_{\rm CR}$ of extragalactic UHECRs is of the order
of  a few 10 Mpc. It cannot be arbitrarily large for protons above
the Greisen-Zatsepin-Kuzmin cutoff of $5\cdot 10^{19}$ eV (Greisen
1966, Zatsepin \& Kuzmin 1966) due to strong energy losses by
photo-pion production in interactions with the CMB. Therefore the
observed UHECRs are believed to be produced within the local
supercluster (LSC). We want to examine the effect of a population of
ghosts within the IGM of the LSC on the propagation of UHECRs. 
Note that we momentarily ignore any magnetic fields in the space
between ghosts, which are extensively discussed in the literature (see
e.g. Sigl et al.  (1999), or Medina-Tanco (1998), and references
therein).

We assume that a small fraction $\Phi_{\rm Vol}$ of the volume of the
LSC is filled with ghosts, which we assume to have a single, typical
size of $l_{\rm ghost}$. The number density of ghosts is therefore $
n_{\rm ghost} = \Phi_{\rm Vol} \, l_{\rm ghost}^{-3}$.  The
cross-section for an UHECR particle to hit a ghost is $\sigma_{\rm
  ghost} = l_{\rm ghost}^2$ leading to a mean free path of
\begin{eqnarray}\nonumber
  L_{\rm ghost} &=& \frac{1}{\sigma_{\rm ghost}\,n_{\rm ghost}} =
  10 \,{\rm Mpc}\, \left(\frac{ l_{\rm ghost} }{\rm 0.1 \,Mpc} \right)
  \left( \frac{\Phi_{\rm Vol}}{0.01} \right)^{-1}.
\end{eqnarray}

Let us assume that a UHECR particle of energy $E_{\rm CR}$ entered a ghost.
Its propagation direction will be deflected due to gyration in the
magnetic fields. We assume that the magnetic fields are coherent on a
scale $l_{\rm cell}$, so that the typical change in direction during
the passage through one coherence length is the angle 
\begin{equation}
  \label{eq:cell}
   \alpha_{\rm cell} = l_{\rm cell}/r_{\rm g}\;,
\end{equation}
where
\begin{equation}
  \label{eq:rg}
  r_{\rm g} = \frac{E_{\rm CR}}{e\,B} = 0.1 \,{\rm Mpc}\, \left(
      \frac{E_{\rm CR}}{\rm 10^{20}\, eV} \right) \, \left(
      \frac{B}{\mu \rm G} \right)^{-1}
\end{equation}
is the gyro-radius of the particle. The accumulated angle change during
the passage through the ghosts by $N_{\rm cell} = l_{\rm ghost}/l_{\rm
  cell}$ regions of random field orientations is given by random
walk arguments to be
\begin{equation}
  \label{eq:alpha.ghost}
  < \alpha_{\rm ghost}^2 > \, = N_{\rm cell} \, 
  \alpha_{\rm cell}^2 = l_{\rm ghost} \,l_{\rm cell}/r_{\rm g}^2\;.
\end{equation}
Isotropic scattering occurs if $\sqrt{<\alpha_{\rm ghost}^2>} \ge
\pi/2$, or equivalently 
\begin{eqnarray}\nonumber
  \label{eq:totalscatt}
  E_{\rm CR} &\le& E_* = 2\cdot 10^{19}\,{\rm eV} \,\left(
    \frac{B}{\mu \rm G} \right) \, \left( \frac{l_{\rm ghost}\,l_{\rm
        cell}}{\rm 0.1\, Mpc \times 0.01\, Mpc} \right)^{\frac{1}{2}}. 
\end{eqnarray}
Particles above $E_*$ experience only a small angle scattering during
the passage through a ghost. The accumulated effect of several ghosts
($N_{\rm ghost}= L_{\rm CR}/L_{\rm ghost}$) along the path of length
$L_{\rm CR}$ of a UHECR is given by
\begin{eqnarray}
  \label{eq:crpath}
  < \alpha^2 > &= & N_{\rm ghost} <\alpha_{\rm ghost}^2 > = \Phi_{\rm Vol}
  \,L_{\rm CR} \, l_{\rm cell} /r_{\rm g}^2 \\ 
  \sqrt{< \alpha^2 >} &=& 18^\circ \, \, \left( \frac{E_{\rm
        CR}}{10^{20}\, \rm eV} \right)^{-1} \left( \frac{B}{\mu \rm G}
  \right) \, \times \nonumber \\
&&\left( \frac{L_{\rm CR}\,l_{\rm cell}}{\rm 10\, Mpc
      \times 0.01\, Mpc} \right)^{\frac{1}{2}}\, \left(
    \frac{\Phi_{\rm Vol}}{0.01} \right)
\end{eqnarray}
This demonstrates that the propagation of UHECRs can be strongly
affected by even a small filling factor of ghosts ($\Phi_{\rm Vol} =
0.01$). The arrival directions are changed by some random angle for
UHECRs above a typical energy $E_*$. UHECRs below this energy are
isotropically scattered by ghosts, so that the length of their
trajectories can be much longer than the distance to the source. For
particles above the Greisen-Zatsepin-Kuzmin cutoff this can have
drastic effects on the observed UHECR spectrum, since the consequent
prolongation of the path reduces the flux due to the corresponding
higher photo-pion energy losses (Medina-Tanco \& En{\ss}lin 1999).

\section{Conclusion}

Radio ghosts, patches of old, remnant, invisible radio plasma, are
expected to be found frequently in the denser regions of the Universe
due to the strong quasar activity at earlier epochs. Their direct
detection is very difficult. But indirect methods like radio
observations of re-activated electron populations in shocked ghosts
seemed to have revealed them nearly 2 decades ago in the cluster radio
relic phenomena. Their magnetic fields can can deflect ultra high
energy cosmic rays significantly in the course of their propagation
through intergalactic space. Further, the fields should trap the
relativistic particle population of former radio galaxies. They
should also be able to release a significant number of trapped particles
under the influence of strong turbulence. The latter effect is
proposed as a possible proton injection mechanism into the
intra-cluster medium leading to radio halos via hadronic secondary
electron and mainly positron production.

We conclude that, although the existence of ghosts is
speculative, their important influence on the physics of the
intergalactic medium requires further investigations.


\invis{
\begin{acknowledgements}
  I gratefully acknowledge many interesting discussions about the
  different aspects of radio ghosts with Peter Biermann, Tracy Clarke,
  Quentin Dufton, Phil Kronberg, and Gustavo Medina-Tanco, who
  invented the short term `ghosts' for my `invisible patches of old,
  remnant radio plasma'. This research was supported by the {\it
    National Science and Engeneering Research Council of Canada}
  (NSERC) and the {\it Max-Planck-Institut f\"ur Radioastronomie}
  (MPIfR-Bonn).
\end{acknowledgements}
   }

\end{document}